\documentclass[superscriptaddress,preprint,amsmath,amssymb,aps,showkeys,prb]{revtex4-2}

\usepackage{graphicx}
\usepackage{dcolumn}
\usepackage{bm}
\usepackage[colorlinks=true,linkcolor=blue]{hyperref}

\begin{document}
	
	\title{Plasmonic nanocavity-enabled universal detection of layer-breathing vibrations in two-dimensional materials}
	\author{Heng Wu}
	\thanks{These authors contributed equally to this work.}
	\affiliation{State Key Laboratory of Semiconductor Physics and Chip Technologies, Institute of Semiconductors, Chinese Academy of Sciences, Beijing 100083, China}
	\affiliation{Center of Materials Science and Optoelectronics Engineering, University of Chinese Academy of Sciences, Beijing 100049, China}
	\author{Miao-Ling Lin}
	\thanks{These authors contributed equally to this work.}
	\affiliation{State Key Laboratory of Semiconductor Physics and Chip Technologies, Institute of Semiconductors, Chinese Academy of Sciences, Beijing 100083, China}
	\affiliation{Center of Materials Science and Optoelectronics Engineering, University of Chinese Academy of Sciences, Beijing 100049, China}
	\author{Sen Yan}
	\affiliation{State Key Laboratory of Physical Chemistry of Solid Surfaces, College of Chemistry and Chemical Engineering, Xiamen University, Xiamen 361005, China}
	\author{Lin-Shang Chen}
	\affiliation{State Key Laboratory of Semiconductor Physics and Chip Technologies, Institute of Semiconductors, Chinese Academy of Sciences, Beijing 100083, China}
	\author{Zhong-Jie Wang}
	\affiliation{State Key Laboratory of Semiconductor Physics and Chip Technologies, Institute of Semiconductors, Chinese Academy of Sciences, Beijing 100083, China}
	\affiliation{Center of Materials Science and Optoelectronics Engineering, University of Chinese Academy of Sciences, Beijing 100049, China}
	\author{Yi-Fei Zhang}
	\affiliation{State Key Laboratory of Semiconductor Physics and Chip Technologies, Institute of Semiconductors, Chinese Academy of Sciences, Beijing 100083, China}
	\affiliation{Center of Materials Science and Optoelectronics Engineering, University of Chinese Academy of Sciences, Beijing 100049, China}
	\author{Ti-Ying Zhu}
	\affiliation{State Key Laboratory of Semiconductor Physics and Chip Technologies, Institute of Semiconductors, Chinese Academy of Sciences, Beijing 100083, China}
	\affiliation{Center of Materials Science and Optoelectronics Engineering, University of Chinese Academy of Sciences, Beijing 100049, China}
	\author{Zheng-Yu Su}
	\affiliation{State Key Laboratory of Semiconductor Physics and Chip Technologies, Institute of Semiconductors, Chinese Academy of Sciences, Beijing 100083, China}
	\author{Jun Wang}
	\affiliation{State Key Laboratory of Physical Chemistry of Solid Surfaces, College of Chemistry and Chemical Engineering, Xiamen University, Xiamen 361005, China}
	\author{Xue-Lu Liu}
	\affiliation{State Key Laboratory of Semiconductor Physics and Chip Technologies, Institute of Semiconductors, Chinese Academy of Sciences, Beijing 100083, China}
	\author{Zhong-Ming Wei}
	\affiliation{State Key Laboratory of Semiconductor Physics and Chip Technologies, Institute of Semiconductors, Chinese Academy of Sciences, Beijing 100083, China}
	\affiliation{Center of Materials Science and Optoelectronics Engineering, University of Chinese Academy of Sciences, Beijing 100049, China}
	\author{Yan-Meng Shi}
	\affiliation{State Key Laboratory of Semiconductor Physics and Chip Technologies, Institute of Semiconductors, Chinese Academy of Sciences, Beijing 100083, China}
	\affiliation{Center of Materials Science and Optoelectronics Engineering, University of Chinese Academy of Sciences, Beijing 100049, China}
	\author{Xiang Wang}
	\affiliation{State Key Laboratory of Physical Chemistry of Solid Surfaces, College of Chemistry and Chemical Engineering, Xiamen University, Xiamen 361005, China}
	\author{Bin Ren}
	\affiliation{State Key Laboratory of Physical Chemistry of Solid Surfaces, College of Chemistry and Chemical Engineering, Xiamen University, Xiamen 361005, China}
	\author{Ping-Heng Tan}
	\email{phtan@semi.ac.cn}
	\affiliation{State Key Laboratory of Semiconductor Physics and Chip Technologies, Institute of Semiconductors, Chinese Academy of Sciences, Beijing 100083, China}
	\affiliation{Center of Materials Science and Optoelectronics Engineering, University of Chinese Academy of Sciences, Beijing 100049, China}
	
	\begin{abstract}
		Conventional Raman spectroscopy faces inherent limitations in detecting interlayer layer-breathing (LB) vibrations with inherently weak electron-phonon coupling or Raman inactivity in two-dimensional materials, hindering insights into interfacial coupling and stacking dynamics. Here we demonstrate a universal plasmon-enhanced Raman spectroscopy strategy using gold or silver nanocavities to strongly enhance and detect LB modes in multilayer graphene, hBN, and their van der Waals heterostructures. Plasmonic nanocavities even modify the linear and circular polarization selection rules of the LB vibrations. By developing an electric-field-modulated interlayer bond polarizability model, we quantitatively explain the observed intensity profiles and reveal the synergistic roles of localized plasmonic field enhancement and interfacial polarizability modulation. This model successfully describes the behavior across different material systems and nanocavity geometries. This work not only overcomes traditional detection barriers but also provides a quantitative framework for probing interlayer interactions, offering a versatile platform for investigating hidden interfacial phonons and advancing the characterization of layered quantum materials.
	\end{abstract}
	
	\maketitle
	
	\section{Introduction}
	The strategic design of metallic nanocavities enables localized surface plasmon resonance to confine the optical field into sub-wavelength volumes, achieving extraordinary local electromagnetic field enhancement by several orders of magnitude\cite{schuller2010NatMater,giannini2011ChemRev,halas2011ChemRev,ding2016NRM}. This provides unprecedented opportunities for plasmon-enhanced Raman spectroscopy (PERS) to uncover light-matter interactions at the nanoscale\cite{wen2022LSA,li2024AM}. The technique leverages both electromagnetic and chemical enhancement mechanisms to boost the Raman signal of molecular vibration modes, achieving single-molecule detection sensitivity\cite{kneipp1997PRL,nie1997Science,xu1999PRL,zhang2013Nature,wang2020NRP,itoh2023ChemRev,yi2025CSR}. In addition, plasmonic nanocavities spatially reconfigure the incident optical field, modifying Raman selection rules and activating symmetry-forbidden modes in conventional Raman spectroscopy\cite{ikeda2013JACS,chen2018LSA,lu2024NC}. These unique characteristics render PERS an indispensable tool for surface and interface analysis at the molecular level, as well as for in situ characterization of nanomaterials\cite{willets2007ARPC,ding2016NRM,wang2020NRP,itoh2023ChemRev}.
	
	PERS further advances studies on the fundamental properties of two-dimensional materials (2DMs) by leveraging their atomically thin nature, which enables exceptionally efficient coupling with plasmonic fields. Among them, the remarkable enhancement and selective excitation of intralayer vibration modes in various 2DMs\cite{sheng2017PRL,chen2018LSA,lu2024NC} have enabled precise characterization of localized lattice dynamics\cite{gadelha2021Nature}, edge-related properties\cite{huang2019NC}, and quantitative measurements of the plasmonic field enhancement in nanocavities\cite{chen2018LSA,lu2024NC}. A promising application of PERS lies in the investigation of low-frequency interlayer modes\cite{tan2012NM} in 2DMs. The inherent Raman inactivity or inherently weak electron-phonon coupling (EPC) of some interlayer modes prevents their observation via conventional Raman spectroscopy\cite{tan2012NM,stenger20172DM}. Remarkably, the out-of-plane displacement fields of interlayer phonon modes exhibit a spatial extent comparable to subwavelength plasmonic near-fields, in stark contrast to the atomic-scale confinement of intralayer modes. This unique overlap suggests significant modifications in the PERS response, a currently underexplored phenomenon requiring systematic study.
	
	In this study, we demonstrated significant enhancement of interlayer layer-breathing (LB) modes in multilayer graphene, hBN and their van der Waals heterostructures (vdWHs) coupled with plasmonic gold (AuNCs) or silver (AgNCs) nanocavities. Upon resonant excitation, the AuNCs and AgNCs generate strong localized plasmonic fields and induce polarizability modifications in 2DMs. These synergistic effects enhance the LB mode signals and modulate the intensity profile of the observed LB modes in AuNC- or AgNC-coupled multilayer graphene, hBN and related vdWHs. We developed an electric-field-modulated interlayer bond polarizability model (E-IBPM) that comprehensively accounts for both variations in interlayer bond polarizability and the inhomogeneity of the plasmonic field within 2DMs. This framework successfully explains the observed Raman intensity profile of LB modes in multilayer graphene, hBN and the related vdWHs coupled with AuNCs or AgNCs, a finding bridges the gap between the localized plasmonic enhancement and the Raman intensity of interlayer modes. The proposed methodology promises to be applicable to other elusive quasiparticles, such as interlayer excitons and certain plasmonic resonances in vdWHs.
	
	\section{Results}
	\subsection{Plasmon-enhanced Raman modes in AuNCs/\texorpdfstring{$N$}{}LG}
	
	In conventional Raman scattering of phonons, the Raman intensity is critically governed by the EPC strength. This renders interlayer phonons in 2DMs (e.g., the LB modes in multilayer graphene) with weak EPC strength particularly challenging to detect. PERS can overcome this longstanding limitation, as it can amplify weak Raman signals by simultaneously enhancing the electric field of incident/scattered light and modulating the polarizability of the materials. Motivated by these advantages, we employed PERS to characterize the LB modes of multilayer graphene.
	
	\begin{figure*}[htb!]
		\includegraphics[width=\linewidth]{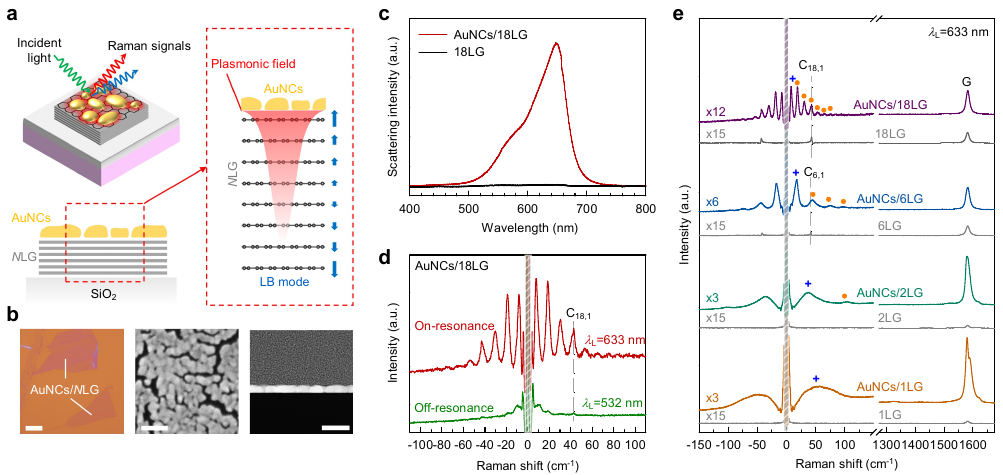}
		\caption{\label{fig1} \textbf{Plasmon-enhanced Raman modes in AuNCs/$N$LG.} (a) Schematic illustration of the AuNCs/$N$LG and the PERS of LB modes. (b) Optical (left panel, scale bar is 20 $\mu$m), SEM (middle panel, scale bar is 40 nm), and TEM (right panel, scale bar is 25 nm) images of AuNCs/$N$LG. (c) Dark-field scattering spectra of AuNCs/18LG and 18LG. (d) Low-frequency Raman spectra of AuNCs/18LG under on-resonance ($\lambda_{\text{L}}=633$ nm) and off-resonance ($\lambda_{\text{L}}=532$ nm) conditions. (e) Raman spectra of AuNCs/$N$LG and the corresponding $N$LG with $\lambda_{\text{L}}$ of 633 nm ($N=1,2,6,18$). The "*" and "+" symbols indicate the LB modes and new low-frequency modes, respectively. }
	\end{figure*}
	
	Figure \ref{fig1}(a) illustrates the PERS schematic for detecting LB modes in $N$-layer graphene ($N$LG) through coupling with AuNCs. Under the plasmon resonance condition, AuNCs generate a strongly confined plasmonic field localized at the nanoscale around their surfaces to enhance the intensity of the LB modes in $N$LG with inherently weak electron-phonon coupling or Raman inactivity. In our experiments, we fabricated AuNCs/$N$LG structures by depositing an 8 nm Au film on $N$LG. Notably, the $N$ of the $N$LG flake was pre-determined using Raman spectroscopy\cite{Li-2015-Nanoscale} prior to Au deposition. The Au nanofilm comprises nanoislands which are varied in size and irregular in shape, separated by numerous nanogaps, as elucidated by the scanning electron microscopy (SEM) and transmission electron microscopy (TEM) images (Fig.\ref{fig1}(b)), which have been proven to serve as AuNCs with good plasmonic activity\cite{lee2011ACSNano,zhao2014Nanoscale}.
	Indeed, the dark-field scattering spectra of AuNCs/18LG (Fig.\ref{fig1}(c)) suggest a plasmon resonance peak at around 633 nm. When performing low-frequency Raman measurements on AuNCs/18LG under both plasmon on-resonance ($\lambda_{\text{L}}=633$ nm) and off-resonance ($\lambda_{\text{L}}=532$ nm) excitation conditions, we found significant enhancement of Raman modes under on-resonance excitation compared to the off-resonance case (Fig.\ref{fig1}(d)), unambiguously confirming the critical role of plasmon resonance in amplifying the Raman mode signals.
	
	To confirm the robustness of the enhanced Raman modes in AuNCs/$N$LG, we further acquired the Raman spectra of AuNCs/$N$LG and the corresponding pristine $N$LG ($N=1,2,6,18$) excited by a laser wavelength ($\lambda_{\text{L}}$) of 633 nm, as illustrated in Fig.\ref{fig1}(e). The G mode ($\sim 1580$ cm$^{-1}$) of AuNCs/$N$LG exhibits significant enhancement compared with that of $N$LG when $N=1,2$, and exhibits obvious strain-induced mode splitting when $N=1$. As $N$ increases, the enhancement effect weakens and the mode splitting vanishes, which is consistent with previous results and can be attributed to the weaker interfacial interaction between AuNCs and $N$LG for large $N$\cite{lee2011ACSNano}. The lack of a D mode ($\sim 1360$ cm$^{-1}$) in the Raman spectra (Fig.\ref{fig1}(e)) indicates minimal defects\cite{jorio2010raman} in the AuNCs/$N$LG fabrication process. This validates the use of the pre-determined $N$ for constructing physical models to reliably interpret the measured plasmon-enhanced Raman spectra in AuNCs/$N$LG.
	
	Beyond the high-frequency intralayer modes, the distinct low-frequency interlayer modes of $N$LG also warranted particular attention\cite{tan2012NM}. Theoretically, $N$LG has $N$-1 pairs of doubly degenerate C modes and $N$-1 LB modes, which can be denoted by C$_{N,N-j}$ and LB$_{N,N-j}$ ($j=1,2,\cdots,N-1$) modes, respectively. The C$_{N,1}$ (LB$_{N,1}$) mode represents the C (LB) mode with the highest frequency. After subtracting the background signals from AuNCs or substrates, clear low-frequency Raman features emerge in both AuNCs/$N$LG and $N$LG (see Supplementary Note 1 and Fig.S1). Figure \ref{fig1}(e) demonstrates that all LB modes remain absent in $N$LG with $N=2,6,18$ due to their exceptionally weak EPC, while weak C$_{N,1}$ modes can be observed. Nonetheless, the C$_{N,1}$ modes are not evident in AuNCs/$N$LG. Instead, a series of layer-number-dependent Raman features emerged below 130 cm$^{-1}$, as marked by asterisks and crosses in Fig.\ref{fig1}(e). The absence of C$_{N,1}$ modes in AuNCs/$N$LG can be attributed to the strong spectral overlap between these new modes and the inherently weak C$_{N,1}$ modes, combined with the significantly higher intensity of the new modes. Notably, the frequencies of certain new modes in AuNCs/$N$LG surpass those of the C$_{N,1}$ modes, i.e., the highest-frequency C modes in pristine $N$LG\cite{tan2012NM}. Based on this evidence, these new modes are preliminarily assigned to plasmon-enhanced LB modes of AuNCs/$N$LG. Raman mapping of a typical AuNCs/16LG sample in Fig.S2 further confirmed the scalability and reproducibility of the PERS-based method for enhancing LB modes in $N$LG. Notably, there are subtle differences between the plasmon-enhanced LB modes of AuNCs/$N$LG and the expected LB modes in pristine $N$LG. For instance, a new low-frequency mode is observable in AuNCs/1LG, whereas no LB mode is expected in pristine 1LG. These distinctions will be further elaborated in subsequent sections.
	
	\subsection{Plasmon-enhanced LB modes in AuNCs/\texorpdfstring{$N$}{}LG}
\begin{figure*}[htb!]
		\includegraphics[width=\linewidth]{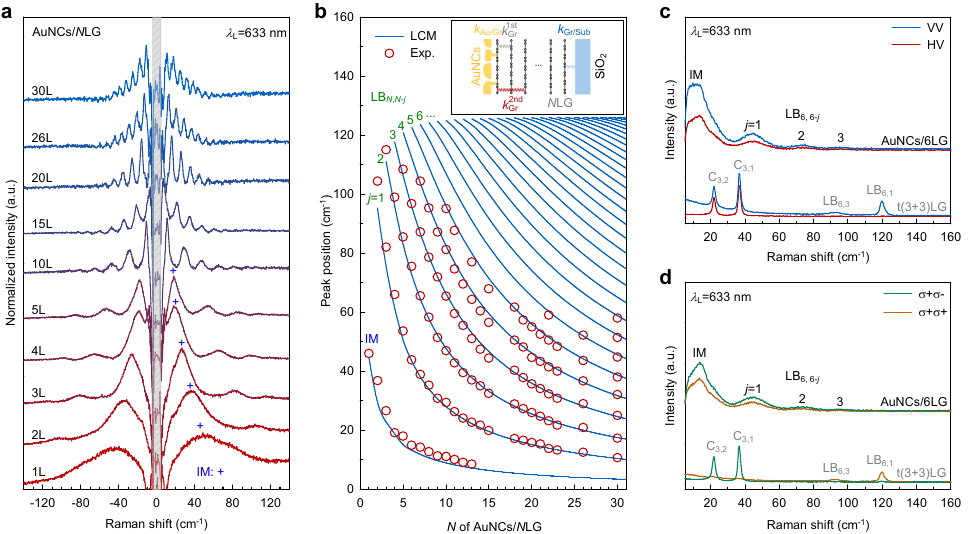}
		\caption{\label{fig2} \textbf{Plasmon-enhanced LB modes in AuNCs/$N$LG.} (a) Low-frequency Raman spectra (up to 140 cm$^{-1}$) of AuNCs/$N$LG, where $N$ varies from 1 to 30. The IMs are marked by "+". (b) Extracted LB mode frequencies from (a) as a function of $N$ (circles), along with the fitting results (lines) by the LCM. The inset illustrates the LCM for AuNCs/$N$LG and the corresponding LB force constants. (c) Linearly polarized (including VV and HV configurations) and (d) circularly polarized (including $\sigma+\sigma+$ and $\sigma+\sigma-$ configurations) Raman spectra of AuNCs/6LG and t(3+3)LG ($\theta$=10.5$^{\circ}$). The C and LB modes are labeled. All spectra were obtained with $\lambda_{\text{L}}=633$ nm.}
	\end{figure*}

	To elucidate the physical origins of plasmon-enhanced LB modes in  AuNCs/$N$LG, systematic low-frequency Raman spectroscopic investigations (up to 140 cm$^{-1}$) were conducted on AuNCs/$N$LG systems ($1\leq N\leq30$) using $\lambda_{\text{L}}$=633 nm excitation. The background-subtracted spectra are depicted in Fig.\ref{fig2}(a), while Fig.\ref{fig2}(b) summarizes the frequencies of the observed LB modes, revealing a pronounced $N$-dependence.
	
	To better understand these LB mode frequencies, the linear chain model (LCM)\cite{tan2012NM} was employed for AuNCs/$N$LG, which treats $N$-layer graphene as a system of rigid masses connected by springs\cite{tan2012NM,wu2022npj2DMA}. The rigid mass represents the mass per unit area of each graphene layer, and the force constants of springs correspond to the interlayer coupling strength between adjacent graphene layers. However, we observed $N$ LB modes in AuNCs/$N$LG ($N=1,2,3$), deviating from the established theoretical LCM framework that predicts $(N-1)$ LB modes in pristine $N$LG ($N>1$). We note that the robust interfacial coupling at both the graphene/substrate and AuNCs/graphene interfaces would perturb the out-of-plane acoustic phonon, causing it to emerge at frequencies lower than the expected LB modes. This perturbed mode, referred to as the interface mode (IM) of the hybrid structure, has been observed in various 2DMs like CVD-grown Bi$_{2}$Te$_{3}$ and WS$_{2}$\cite{zhao2014PRB,xiang2021NanoRes}. A detailed comparison of frequencies and interlayer displacements for the out-of-plane acoustic modes and LB modes in $N$LG along with the LB modes and IM in AuNCs/$N$LG is provided in Fig.S3. Thus, we modified the LCM to incorporate the interfacial LB force constants (i.e., $k_{\text{Gr/Sub}}$ at the graphene/substrate interface and $k_{\text{Au/Gr}}$ at the AuNCs/graphene interface) at the two interfaces, in addition to the nearest-neighbor interlayer LB force constant in $N$LG ($k_{\text{Gr}}^{\text{1st}}=10.7\times10^{19}$ N$\cdot$m$^{-3}$) and the next-nearest-neighbor LB force constant ($k_{\text{Gr}}^{\text{2nd}}=0.9\times10^{19}$ N$\cdot$m$^{-3}$)\cite{wu2015ACSNano}. This extended model successfully describes both IM and LB mode frequencies. By fitting the extracted LB mode frequencies in Fig.\ref{fig2}(b) with the modified LCM, we obtained $k_{\text{Gr/Sub}}=0.2k_{\text{Gr}}^{\text{1st}}$ and $k_{\text{Au/Gr}}=0.3k_{\text{Gr}}^{\text{1st}}$ (See Supplementary Note 2 in SI for the full description of LCM). The excellent agreement between theoretical and experimental results in Fig.\ref{fig2}(b) underscores the essential role of interfacial coupling in properly understanding the LB mode frequency of AuNCs/$N$LG probed by PERS.
	
	PERS follows fundamentally different selection rules compared to conventional Raman scattering, providing a clear spectroscopic feature to distinguish these two mechanisms\cite{leru2009a,itoh2023ChemRev}. This distinction is exemplified by comparing the polarization-dependent responses of twisted (folded) double trilayer graphene (t(3+3)LG) and AuNCs/6LG systems. In t(3+3)LG with a twist angle of 10.5$^{\circ}$, the C$_{3,1}$ and C$_{3,2}$ modes of the 3LG constituent and the LB$_{6,1}$ and LB$_{6,3}$ modes of 6LG are observed. This distinct difference between the C and LB modes stems from the contrasting interfacial coupling in twisted multilayer graphene (tMLG): while interfacial shear coupling is very weak, the LB coupling strength approaches that of interlayer coupling in graphite and AB-stacked multilayer graphenes\cite{wu2015ACSNano}. Consequently, the C mode's layer displacements are localized within the corresponding constituent, whereas the LB mode involves collective vibrations across all stacked layers in twisted (folded) multilayer graphenes\cite{wu2015ACSNano}. This behavior is a general characteristic of C and LB modes in vdWHs\cite{liang2017ACSNano,li2017ACSNano,lin2019NC,hao2023ACSNano}, providing a diagnostic criterion to distinguish them in experimental observations once the thicknesses of all the constituents are known. The two modes of t(3+3)LG exhibit strong enhancement under 633 nm excitation, in resonance with the twist-angle-dependent van Hove singularity (VHS) transition\cite{wu2014NC,wu2015ACSNano}. However, they display markedly different polarized responses\cite{wu2015ACSNano}: the C mode exhibits non-zero intensity in linearly parallel-polarized (VV), linearly crossed (HV), and circularly cross-polarized ($\sigma+\sigma-$) configurations, while the LB modes show remarkable intensity in VV and circular co-polarized ($\sigma+\sigma+$) polarizations. These observations perfectly align with conventional Raman selection rules\cite{wu2015ACSNano,chen2015NL}. In striking contrast, the LB mode intensity in AuNCs/6LG remains non-zero across all four polarization configurations while showing negligible polarization dependence. This suggests that the AuNCs do not just passively enhance the signal by plasmonic field enhancement, but actively modify the Raman tensor properties associated with the bond polarizability of 6LG. These observations unambiguously demonstrate that AuNCs substantially altered the interlayer bond polarizability properties and the spatial distribution of the incident electric field in the Raman spectroscopy of adjacent multilayer graphene.
	
	\subsection{Electric-field-modulated interlayer bond polarizability model}
	
	\begin{figure}[htb!]
		\includegraphics[width=\linewidth]{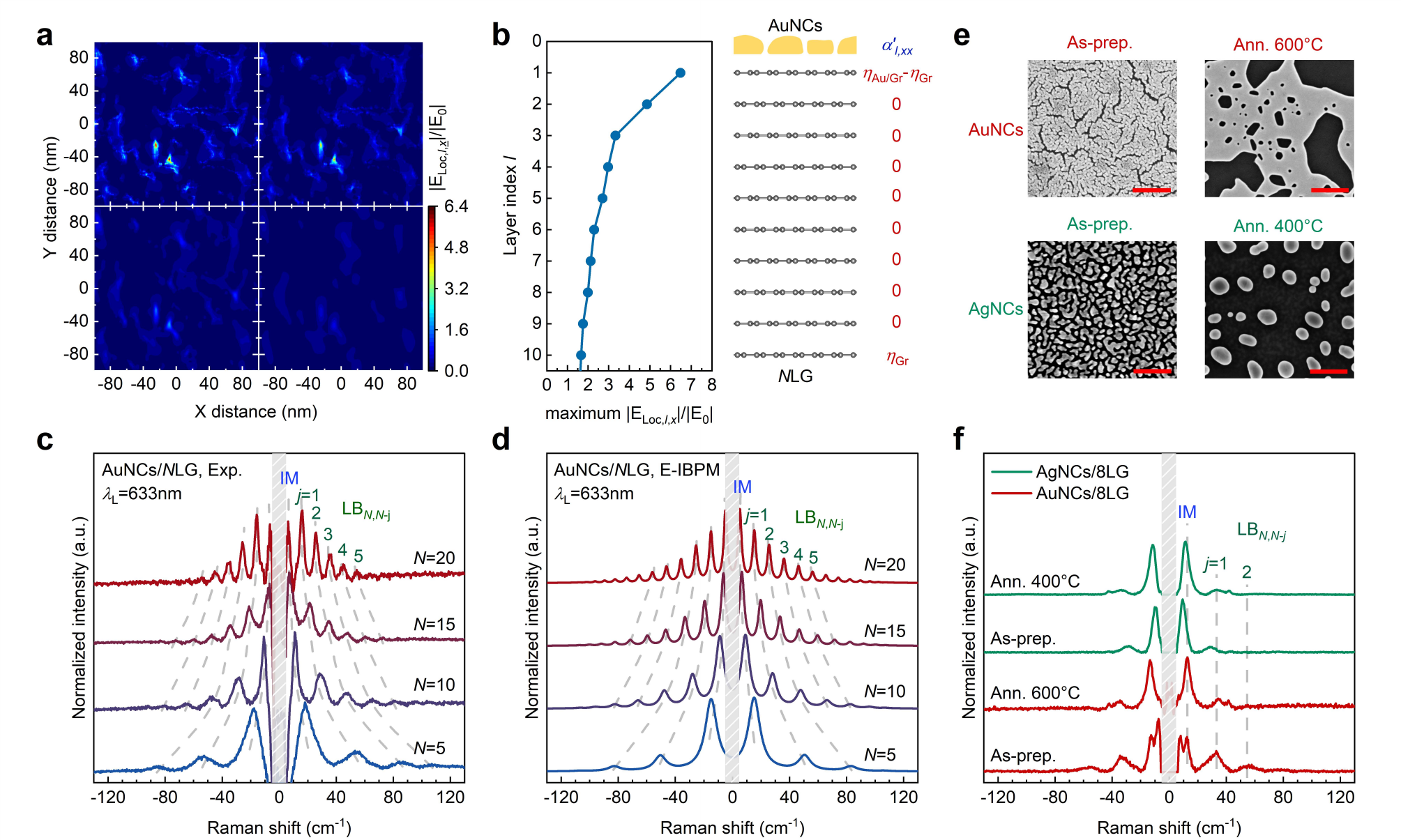}
		\caption{\label{fig3} \textbf{Electric-field-modulated interlayer bond polarizability model for plasmon-enhanced LB modes in AuNCs/$N$LG.} (a)  FDTD-simulated distribution of normalized $x$-component of local field ($|E_{\text{Loc},l,x}|/|E_0|$) in different graphene layers near AuNCs. (b) Simulated maximum $|E_{\text{Loc},l,x}|/|E_0|$ at graphene layers near AuNCs and a schematic of AuNCs/$N$LG with $\alpha_{i,xx}^{\prime}$. (c) Measured PERS spectra with $\lambda_{\text{L}}=633$ nm and (d) the calculated IM and LB modesof AuNCs/$N$LG ($N$=5, 10, 15, 20) by E-IBPM. The IM and LB modes are labeled. The lineshapes of all LB modes were modeled as Lorentzian functions, with FWHM of 10 cm$^{-1}$, 7 cm$^{-1}$, 4 cm$^{-1}$ and 3 cm$^{-1}$ for $N$=5, 10, 15 and 20, respectively. (e) SEM images and the (f) corresponding Raman spectra of as-prepared AuNCs/8LG, AuNCs/8LG annealed at 600$^\circ$C, as-prepared AgNCs/8LG and AgNCs/8LG annealed at 400$^\circ$C. Scale bars in (e): 250 nm. The IMs and LB modes are marked with dashed lines in (f).}
	\end{figure}
	
	The intensity analysis of Raman modes in 2DMs provides deeper insights into their underlying scattering mechanisms. It is noteworthy that quantitatively comparing the absolute Raman intensity of a specific mode across different samples is inherently challenging, as it is influenced by a multitude of factors including plasmonic field distribution, optical interference effects, and experimental conditions. Therefore, the relative intensity between the LB modes within the same spectrum often provides a more robust and physically insightful metric. For the interlayer modes in pristine 2DMs, the interlayer bond polarizability model (IBPM)\cite{liang2017Nanoscale} well explains their relative intensity by computing the Raman tensors from polarizability changes induced by interlayer bonds between adjacent layers. Within the electric dipole approximation, the IBPM neglects spatial variations in the incident electric field, linking the Raman intensity of interlayer modes to the derivative of the polarizability tensor along layer displacements. In contrast, the PERS of LB modes in AuNCs/2DM systems violates the electric dipole approximation due to a strongly confined plasmonic field with nanometer-scale spatial localization at the AuNCs/2DM interface. In addition, the  interfacial interaction between 2DMs and AuNCs\cite{wang2020NRP,itoh2023ChemRev,li2024AM} can modify the polarizability of the 2DMs. Thus, a new theoretical framework incorporating both local field enhancement and polarizability modulation is required to accurately model the plasmon-enhanced intensity profile of the LB modes in AuNCs/2DM systems.
	
	In PERS, the Raman response of a plasmon-enhanced mode can be characterized by the induced Raman dipole moment $\mathbf{p}$, closely linked to its Raman polarizability tensor $\Delta\boldsymbol{\alpha}_{j}$ and the localized plasmonic field generated by AuNCs, i.e., $\mathbf{p}=\Delta\boldsymbol{\alpha}_{j}\cdot\mathbf{E}_{\text{Loc}}$. Unlike free-space dipole emission, the radiation from this dipole near the metal nanostructure is strongly modified by the plasmonic effects. Both its emission properties and the local excitation field are enhanced through analogous mechanisms. The far-field detectable Raman signal is thus determined by this modified radiation. The radiation process is quantitatively described by the Green's function $\mathbf{G}$, which relates the self-reaction radiation field ($\mathbf{E}_{\text{R}}$) to the Raman dipole moment, $\mathbf{E}_{\text{R}}=\mathbf{G}\cdot\mathbf{p}=\mathbf{G}\cdot\Delta\boldsymbol{\alpha}_{j}\cdot\mathbf{E}_{\text{Loc}}$\cite{leru2009a}. Thus, the PERS intensity of the $j$-th Raman mode in 2DMs can be expressed as\cite{leru2009a,liang2017Nanoscale}
	
	\begin{equation}\label{Placzek}	I_{j}\propto\frac{n_{j}+1}{\omega_{j}}\left| \mathbf{G}\cdot\Delta\boldsymbol{\alpha}_{j}\cdot\mathbf{E}_{\text{Loc}}\right|^{2}
	\end{equation}
	
	\noindent where $\omega_{j}$ is the frequency of the $j$-th mode. $n_{j}=1/[\exp(\hbar\omega_{j}/k_{b}T)-1]$ is the Bose-Einstein factor. The $\mathbf{G}$ is always described by a radiation enhancement factor, depending on several factors, including the substrate geometry and optical properties, the dipole position, orientation, and its emission frequency. Estimating $\mathbf{G}$ is an intractable task. For practical calculations in parallel-polarized backscattering configurations, the nondiagonal elements in $\mathbf{G}$ can be ignored and the diagonal element in $\mathbf{G}$ can be estimated as $\mathbf{G}\approx| \mathbf{E}_{\text{Loc}}|/|\mathbf{E}_{0}|$, assuming that the local field amplitude and polarization remain similar at both the laser and Raman frequencies.
	
	According to the empirical bond polarizability model, the Raman polarizability change $\Delta\boldsymbol{\alpha}_{j}$ can be approximated as a summation of polarizability variations induced by individual bonds\cite{guha1996PRB,saito2010JPCM} (see Supplementary Note 3). For interlayer modes in 2DMs, only interlayer bonds contribute to the overall polarizability changes\cite{liang2017Nanoscale}. Thus, for the LB$_{N,N-j}$ mode, $\Delta\boldsymbol{\alpha}_{j}$ is given by $\Delta\boldsymbol{\alpha}_{j}=\sum_{l}\boldsymbol{\alpha}_{l}^{\prime}\Delta z_{j,l}$, where $\boldsymbol{\alpha}_{l}^{\prime}=\left(\partial \boldsymbol{\alpha}_{j}/\partial z_{j,l}\right)_{0}$ denotes the polarizability derivative associated with the $l$-th layer, and $\Delta z_{j,l}$ is the layer displacement of the $l$-th layer in LB$_{N,N-j}$ modes, obtained from LCM calculations. To account for plasmonic enhancement, the localized plasmonic field at each layer ($\mathbf{E}_{\text{Loc},l}$) and the corresponding Green's function ($\mathbf{G}_{l}$) should be considered. Consequently, the PERS intensity of a given LB mode can be expressed as:
	
	\begin{equation}\label{EIBPM}
		I_{j}\propto\frac{n_{j}+1}{\omega_{j}}\left|\sum_{l}\mathbf{G}_{l}\cdot (\boldsymbol{\alpha}_{l}^{\prime}\Delta z_{j,l})\cdot\mathbf{E}_{\text{Loc}, l}\right|^{2}
	\end{equation}
	
	\noindent This expression highlights that the Raman dipoles at each layer contribute to the LB mode intensity, similar to the case of molecules in an inhomogeneous local field\cite{zhang2021JRS}. The above framework constitutes the core idea of IBPM\cite{luo2015SR,liang2017Nanoscale,lin2019NC,mei2024Nanoscale}, and this plasmonic field-adapted version is termed the electric-field-modulated IBPM (E-IBPM).
	
	Although the out-of-plane LB modes can, in principle, be more efficiently enhanced by the $z$-component of plasmonic field ($E_{\text{Loc},l,z}$)\cite{sheng2017PRL,sheng2019ACSNano}, our experimental configuration employs a backscattering geometry $z(xx)\bar{z}$, which selectively probes the $xx$ component of $\Delta\boldsymbol{\alpha}_{j}$ ($\Delta\alpha_{j,xx}$) and couples to the $x$ component of the local field ($E_{\text{Loc},l,x}$) in Eq.\ref{EIBPM}. While the high numerical aperture (NA = 0.9) of the objective may permit partial collection of signals enhanced by $E_{\text{Loc},l,z}$, the observed intensity profiles are dominated by enhancements arising from $E_{\text{Loc},l,x}$. Moreover, the intensity profiles of the $E_{\text{Loc},l,z}$-enhanced signals exhibit spatial distributions that closely resemble those enhanced by $E_{\text{Loc},l,x}$ (see Supplementary Note 4). As $\alpha_{l,xx}^{\prime}$ is only associated with the interlayer bond on both sides of the $l$-th layer, the interlayer bond parameter $\eta_{l,x}$ related to the $x$ component of the interlayer bond is introduced to express $\alpha_{l,xx}^{\prime}$ (see Supplementary Note 3 for detailed derivation). Thus, $\Delta\alpha_{j,xx}$ can be expressed as
	
	\begin{equation}\label{Dalpha1}
		\Delta\alpha_{j,xx}=\sum_{l}\alpha_{l,xx}^{\prime}\Delta z_{j, l}=\sum_{l}(\eta_{l,x}-\eta_{l+1,x})\Delta z_{j,l}
	\end{equation}
	
	\noindent Since $|G_{l}|\approx |E_{\text{Loc},l,x}|/|E_{0}|$, the intensity of LB modes and IMs is given by
	
	\begin{equation}\label{LBMxx}	
		I_{j}\propto\frac{n_{j}+1}{\omega_{j}} \left|\sum_{l }(\eta_{l,x}-\eta_{l+1,x})\Delta z_{j,l}E_{\text{Loc},l,x}^{2}\right|^{2}
	\end{equation}
	
	\noindent The relative intensity of the LB modes is dictated by the relative magnitude of the plasmonic field at each layer (i.e., the decay of the plasmonic field across layers). In contrast, the absolute Raman intensity scales with the field's absolute strength. This core physical insight remains perfectly valid regardless of the plasmonic cavity shapes and densities. Once the local field generated by the plasmonic cavities is known, the relative Raman intensity profile of plasmon-enhanced LB modes can be obtained. The local field generated by the AuNCs with $\lambda_{\text{L}}=633$ nm was simulated by the finite-difference time-domain (FDTD) method. Figure \ref{fig3}(a) presents the simulated distribution of $|E_{\text{Loc},l,x}|/|E_0|$ corresponding to the arrangement of AuNCs in Fig.\ref{fig1}(b). Figure \ref{fig3}(b) demonstrates a rapid decay of the maximum $|E_{\text{Loc},l,x}|/|E_0|$ with increasing layer index $l$.
	
	For AuNCs/$N$LG, only the 1st layer (adjacent to the AuNCs) and the $N$-th layer (adjacent to the substrate) that show asymmetric interlayer bonds exhibit a non-zero polarizability derivative\cite{liang2017Nanoscale} (Fig.\ref{fig3}(b)), i.e., $\alpha_{1,xx}^{\prime}=\eta_{\text{Au/Gr}}-\eta_{\text{Gr}}$ and $\alpha_{N,xx}^{\prime}=\eta_{\text{Gr}}$, where $\eta_{\text{Au/Gr}}$ and $\eta_{\text{Gr}}$ are interlayer bond parameters at AuNCs/graphene interface and within $N$LG, respectively. Interior layers ($l=2$ to $N-1$) show $\alpha_{l,xx}^{\prime}=\eta_{\text{Gr}}-\eta_{\text{Gr}}=0$. Moreover, the strong interaction between AuNCs and $N$LG results in $\eta_{\text{Au/Gr}}\gg\eta_{\text{Gr}}$ under the plasmon resonance condition, similar to the case of hBN layers at the heterointerface in hBN/WS$_{2}$ vdWHs under exciton resonance\cite{lin2019NC}. Therefore, the Eq.\ref{EIBPM} for the LB$_{N,N-j}$ mode and IM in AuNCs/$N$LG can be simplified as
	
	\begin{equation}\label{AuGr}
		I_{j}\propto\frac{n_{j}+1}{\omega_{j}}\left|\Delta z_{j,1}\right|^{2}
	\end{equation}
	
	\noindent Using the interlayer displacements from LCM calculations, we simulated the intensities of the LB$_{N,N-j}$ mode and IM in AuNCs/$N$LG using $\eta_{\text{Au/Gr}}=10\eta_{\text{Gr}}$. Figures \ref{fig3}(c,d) compare the experimental low-frequency spectra and the simulated PERS spectra by E-IBPM for the LB modes in AuNCs/$N$LG ($N=5, 10, 15, 20$), in which the LB modes were modeled with Lorentzian lineshapes, with full width at half maximum (FWHM) values of 4 cm$^{-1}$, 3 cm$^{-1}$, 2 cm$^{-1}$ and 1 cm$^{-1}$ for $N$ = 5, 10, 15 and 20, respectively. Notably, the different FWHMs for $N$LG with varied $N$ were determined according to the corresponding Raman spectra from experiments, as widely adopted in previous studies\cite{MLLin-PRL-2025,liang2017Nanoscale,mei2024Nanoscale}. The close match between the simulated and experimental LB mode intensities confirms that the E-IBPM accurately describes their underlying physics. The applicability of E-IBPM to $N$LG coupled with AuNCs with different shapes or densities is further confirmed, as indicated in Fig.\ref{fig3}(e,f) and Supplementary Note 5. We also prepared plasmonic cavities with different shapes and densities by annealing AuNCs and AgNCs at different temperatures. The AuNCs annealed at 600$^\circ$C exhibit nanoislands separated by larger nanogaps than those of as-prepared AuNCs. Similar to the as-prepared AuNCs, the as-prepared AgNCs comprised nanoislands with varying sizes and irregular shapes. In contrast, the AgNCs annealed at 400$^\circ$C show well-separated nanohemispheres (Fig.\ref{fig3}(e) and Fig.S4). The PERS spectra of 8LG coupled to these different plasmonic nanocavities exhibited pronounced plasmon-enhanced LB modes, where the peak positions show a slight difference due to the different interfacial coupling constants at AuNCs/8LG (0.3$k_{\text{Gr}}^{\text{1st}}$) and AgNCs/8LG (0.1$k_{\text{Gr}}^{\text{1st}}$). This difference also leads to distinct relative Raman intensity profiles between AuNCs/8LG and AgNCs/8LG, as discussed in detail in Supplementary Note 5. This indicates that the E-IBPM effectively captures the essential physics occurring within the active plasmonic regions, providing a versatile and efficient framework for understanding the fundamental PERS mechanism in our system.
	
	\subsection{Plasmon-enhanced LB modes in various systems coupled to AuNCs or AgNCs}
	
\begin{figure}[htb!]
		\includegraphics[width=1.0\linewidth]{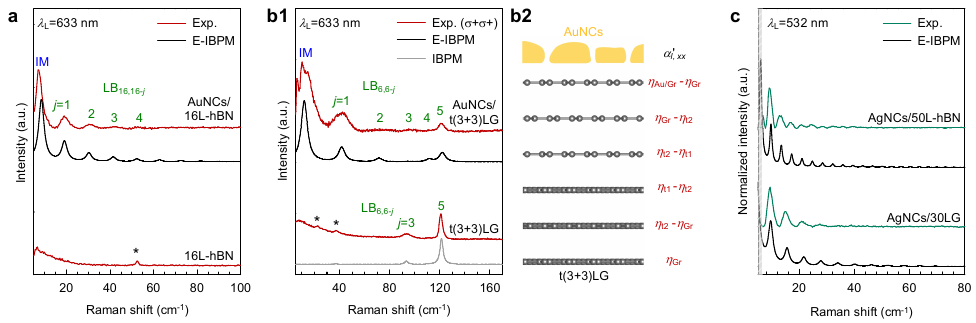}
		\caption{\label{fig4} \textbf{Plasmon-enhanced LB modes in hBN, $N$LG and tMLG coupled to AuNCs or AgNCs.} Experimental (Exp.) and calculated (E-IBPM or IBPM) Raman spectra of the LB modes for (a) AuNCs/16L-hBN, (b1) AuNCs/t(3+3)LG (twist angle $\theta=10.5^{\circ}$) and the corresponding pristine counterparts under 633 nm excitation. The schematic diagram for interlayer bond parameters $\alpha_{l,xx}^{\prime}$ of AuNCs/t(3+3)LG is also shown in (b2). (c) Experimental (Exp.) and calculated (E-IBPM or IBPM) Raman spectra of the LB modes for AgNCs/30LG and AgNCs/50L-hBN under 532 nm excitation. Asterisks (*) indicate residual C modes.}
	\end{figure}

	The generality of the plasmon-enhanced LB mode detection strategy has been further demonstrated in other 2DMs such as hBN, where the extremely weak EPC typically hinders the observation of LB modes\cite{stenger20172DM,lin2019NC}. While the Raman spectrum of pristine 16L-hBN shows only a faint C$_{16,1}$ mode, AuNCs/16L-hBN under resonant excitation ($\lambda_{\text{L}}=633$ nm) exhibits pronounced plasmon-enhanced LB modes after background subtraction (Fig.\ref{fig4}(a)). The assignment of these peaks to LB modes is confirmed by the agreement between experimental frequencies and calculated frequencies using the modified LCM (see Supplementary Note 6). The interlayer LB force constant of multilayer hBN is $k_{\text{hBN}}=9.9\times 10^{19}$ N$\cdot$m$^{-3}$ \cite{wu2022npj2DMA} and the derived interfacial LB force constants are $k_{\text{hBN/Sub}}=2k_{\text{hBN}}$ and $k_{\text{Au/hBN}}=0.1k_{\text{hBN}}$. The significantly different interfacial LB coupling constants compared with those in $N$LG can be ascribed to the different interlayer distance between 2DMs and the underlying substrate (or Au nanofilm) and the mean charge densities at the interfaces\cite{wu2015ACSNano}.
	
	Building on the E-IBPM framework established for AuNCs/$N$LG, we extended the analysis to hBN by considering two critical interfacial parameters, i.e., $\eta_{\text{Au/hBN}}$ (at the AuNCs/hBN interface) and $\eta_{\text{hBN}}$ (within hBN), with $\eta_{\text{Au/hBN}}=10\eta_{\text{hBN}}$. Non-zero polarizability derivatives emerge exclusively in the boundary layers, with $\alpha_{1,xx}^{\prime}=\eta_{\mathrm{Au/hBN}}-\eta_{\mathrm{hBN}}$ for the top layer and $\alpha^{\prime}_{N,xx}=\eta_{\mathrm{hBN}}$ for the bottom layer. The simulated LB mode intensities of AuNCs/$N$L-hBN show excellent agreement with experimental data, where LB modes are modeled with Lorentzian lineshapes featuring a FWHM of 3 cm$^{-1}$ (Fig.\ref{fig4}(a)).
	
	For a complex case, the heterointerface induces non-zero polarizability derivatives for the layers adjacent to the heterointerface, leading to additional contribution to LB mode intensities. This effect is clearly demonstrated in t(3+3)LG (Fig.\ref{fig4}(b)), where the twist-induced electronic VHSs in electronic band structure generate non-zero interior polarizability derivatives $\alpha_{l,xx}^{\prime}$\cite{hao2023ACSNano,mei2024Nanoscale}. Figure \ref{fig4}(b1) compares the Raman spectra of t(3+3)LG and AuNCs/t(3+3)LG ($\theta=10.5^{\circ}$) in $\sigma^+\sigma^+$ configuration measured at $\lambda_{\text{L}}=633$ nm, where the twist angle of t(3+3)LG was identified by the emergent R and R' modes (Fig.S5) in the high-frequency G-band region\cite{Campos-Delgado2013,carozo-prb-2013,Eliel-nc-2018}. In pristine t(3+3)LG, only two high-frequency LB modes (LB$_{6,1}$ and LB$_{6,3}$) are observed. In contrast, in AuNCs/t(3+3)LG, both low- and high-frequency LB modes as well as IM are observed. Remarkably, the frequencies of the observed LB modes can be accurately described using the same LB force constants $k_{\text{Gr}}^{\text{1st}}$, $k_{\text{Gr}}^{\text{2nd}}$ and $k_{\text{Au/Gr}}$ derived from AuNCs/6LG due to the negligible impact of twist interfaces on interlayer couplings\cite{wu2015ACSNano}. The interaction between the substrate and graphene layer is minimized ($k_{\text{Au/Gr}}\approx0$) owing to the specific sample preparation method employed. Consequently, the 1 cm$^{-1}$ blue shift of the LB$_{6,1}$ mode in AuNCs/t(3+3)LG compared with that of t(3+3)LG provides direct evidence for the existence of interfacial coupling at the AuNCs/graphene interfaces.
	
	To simulate the LB mode intensities in AuNCs/t(3+3)LG using E-IBPM, additional interlayer bond parameters are required. Figure \ref{fig4}(b2) illustrates the schematic of AuNCs/t(3+3)LG with the relevant bond parameters: $\eta_{\mathrm{Au/Gr}}$ (AuNCs/graphene interface), $\eta_{\mathrm{t1}}$ (twist interface), $\eta_{\mathrm{t2}}$ (adjacent to twist interface), and $\eta_{\mathrm{Gr}}$ (adjacent to AuNCs/graphene interface). To preserve the structural symmetry of t(3+3)LG, the polarizability derivatives $\alpha_{l,xx}^{\prime}$ are parameterized as follows: $\alpha_{1,xx}^{\prime}=\eta_{\mathrm{Au/Gr}}-\eta_{\mathrm{Gr}}$, $\alpha_{2,xx}^{\prime}=\eta_{\mathrm{Gr}}-\eta_{\mathrm{t2}}$, $\alpha_{3,xx}^{\prime}=\eta_{\mathrm{t2}}-\eta_{\mathrm{t1}}$, and symmetrically for the 4th-6th layers.
	As shown in Fig. \ref{fig4}(b1), the experimental LB mode intensities are well reproduced using parameters $\eta_{\text{Au/Gr}}=1$, $\eta_{\text{t1}}=-1$, $\eta_{\text{t2}}=0.8$, $\eta_{\text{Gr}}=0.1$, combined with the FDTD-simulated $E_{\text{Loc},l,x}$ (Fig. \ref{fig3}(b)). The hierarchy $|\eta_{\mathrm{t1}}|\approx|\eta_{\text{t2}}|\gg|\eta_{\mathrm{Gr}}|$ indicates strong polarizability modulation near the twisted interface during resonance. The negative sign of $\eta_{\mathrm{t1}}$ is attributed to the imaginary part of the polarizability under VHS resonance\cite{liang2017Nanoscale,mei2024Nanoscale}. The validity of these parameters is further supported by the excellent agreement between simulation and experiment for both t(3+3)LG and t(2+2)LG by IBPM and AuNCs/t(2+2)LG by E-IBPM using the same interlayer bond parameters between pristine graphene layers and at twisted interfaces as those in E-IBPM for t(3+3)LG (see Supplementary Note 7 and Fig. S7). Notably, $|\eta_{\mathrm{Au/Gr}}|\gg|\eta_{\mathrm{Gr}}|$ highlights the dramatic plasmon-induced enhancement at the AuNCs interface. This quantitative match across the above distinct material systems (i.e., insulator hBN and tMLG) validates that our theoretical approach correctly captures the essential physics of the PERS of LB modes, implying that the proposed PERS mechanism and E-IBPM are broadly applicable across different AuNCs/2DMs and AuNCs/vdWHs structures, independent of the specific materials.
	
	Furthermore, AgNCs synthesized under varying conditions with different morphologies and densities all exhibit a distinct plasmon resonance peak at 532 nm in their dark-field scattering spectra (see Fig.S11(b)), which suggests the potential to extend the excitation wavelength range of this approach. As examples, Fig.\ref{fig4}(c) presents the observed LB modes in AgNCs/30LG and AgNCs/50L-hBN under 532 nm excitation, along with the corresponding E-IBPM simulation results. For AgNCs/30LG, the frequencies and the relative Raman intensities of the observed plasmon-enhanced LB modes under 532 nm excitation can be accurately described using the same set of LB force constants ($k_{\text{Gr}}^{\text{1st}}$, $k_{\text{Gr}}^{\text{2nd}}$, $k_{\text{Ag/Gr}}$ and $k_{\text{Gr/Sub}}$) and relevant bond parameters ($\eta_{\text{Ag/Gr}}=10\eta_{\text{Gr}}$) as those used under 633 nm excitation. The consistency of these parameters across the two excitation wavelengths indicates strong interactions between AgNCs and $N$LG under both plasmon resonance excitations. For AgNCs/50L-hBN, the frequencies of the plasmon-enhanced LB modes can be well fitted with $k_{\text{hBN}}=9.9\times 10^{19}$ N$\cdot$m$^{-3}$ \cite{wu2022npj2DMA}, $k_{\text{hBN/Sub}}=2k_{\text{hBN}}$ and $k_{\text{Ag/hBN}}=0.02k_{\text{hBN}}$. Remarkably, the relative Raman intensities of these observed LB modes are also well reproduced by the E-IBPM using $\eta_{\text{Ag/hBN}}=10\eta_{\text{hBN}}$, which also suggests a strong interaction between AgNCs and hBN. These results fully validate the feasibility of a wavelength-tunable LB mode detection strategy achieved by utilizing different plasmonic nanocavities. This flexibility allows for the selection of appropriate metal plasmonic cavities for specific material systems, thereby effectively mitigating potential photoluminescence interference.
	
	\section{Discussion}
	In summary, we have developed a PERS approach using AuNCs or AgNCs to characterize LB modes in $N$LG, hBN, and related vdWHs. This technique successfully overcomes the detection limitations imposed by weak EPC in conventional Raman scattering. Combining experimental and theoretical analysis, we elucidate the fundamental enhancement mechanism: the synergistic interplay between interfacial polarization modulation at AuNCs/2DM interface and the localized electric field enhancement of AuNCs dominantly amplifies the Raman dipole moment of the 2DM layer at the interface, contributing to the Raman intensity of the LB modes. The proposed E-IBPM provides a quantitative understanding of the relative LB mode intensities across different Au(Ag)NCs/2DMs and Au(Ag)NCs/vdWHs structures.
	
	Our findings establish: 1) A highly sensitive PERS methodology for probing interlayer phonons and interfacial coupling in 2DMs; 2) A PERS mechanism in 2DMs coupled to AuNCs (AgNCs) and a universal theoretical framework (E-IBPM) for understanding plasmon-coupled interlayer vibrational modes. As interlayer vibration modes carry unique information about the interlayer or interfacial properties in 2DMs and related heterostructures, the plasmonic nanocavity-enabled LB modes in such systems provide a unique, non-destructive tool for characterizing the structure and interlayer coupling at hidden interfaces in complex, multilayer stacks, which are inaccessible to many other techniques. Our work highlights that in complex plasmonic cavity-coupled van der Waals systems, the relative Raman intensity serves as a more reliable probe of the underlying PERS mechanisms than the hard-to-quantify absolute enhancement factors, an insight that is broadly applicable to the field. In addition, the plasmonic nanocavities in this work act as a universal amplifier, capable of not only enhancing intrinsic dipole transitions but also converting weak processes that indirectly couple to the field via polarizability changes into strong, observable signals. The methodology holds potential for application to a broader range of elusive quasiparticles, with interlayer excitons and certain plasmonic resonances in vdWHs being prime candidates. This work opens avenues for precise characterization of interlayer interactions in emerging two-dimensional devices while providing fundamental insights into plasmon-phonon coupling phenomena.
	
	\section{Materials and methods}
	\subsection{Sample preparation}
	Multilayer graphene and hBN flakes were mechanically exfoliated from their bulk crystals and transferred onto 90 nm SiO${_2}$/Si substrates with oxygen plasma-treated surfaces to improve adhesion. The layer number of graphene was determined via Raman spectroscopy by comparing the intensity ratio of the Si peak ($\sim$520.7 cm$^{-1}$) from the substrate beneath the sample to that from an adjacent bare SiO${_2}$/Si substrate under the same focusing condition\cite{Li-2015-Nanoscale}. The thickness of hBN flakes was measured by atomic force microscopy and further verified using the frequencies of the plasmon-enhanced LB modes. The tMLG and hBN/tMLG vdWHs were assembled via a standard polycarbonate-assisted dry-transfer technique. Finally, an 8 nm-thick Au film was thermally evaporated onto the 2DMs and vdWHs at a deposition rate of 5 \AA/s.
	
	\subsection{Structural characterization}
	The surface morphology of the plasmonic nanocavities was characterized by SEM (Helios 5 CX DualBeam, Thermo Scientific) with an accelerating voltage of 3 kV. To investigate the cross-sectional interfacial structure, the samples were prepared by the focused ion beam (FIB) lift-out technique and subsequently analyzed by high-resolution TEM (Talos F200X, Thermo Scientific) at an accelerating voltage of 200 kV.
	
	\subsection{Dark-field scattering spectroscopy}
	The dark-field scattering measurements were performed using a home-made SmartRaman confocal micro-Raman module (Institute of Semiconductors, Chinese Academy of Sciences) coupled
	with a dark-field microscope and a Horiba iHR320 spectrometer. The white light from a tungsten lamp was directed to a $100\times$ objective lens (N.A. = 0.9) to form uniform dark-field illumination. The scattered signal of the samples was collected by the same objective, which was then directed to the iHR320 spectrometer.
	
	\subsection{Raman spectroscopy}
	Raman measurements were performed using a Jobin-Yvon HR800 system equipped with a charge-coupled device (CCD). The Raman scattering signal was collected through a $100\times$ objective lens (N.A. = 0.9). Two laser wavelengths were used in the experiments, including 633 nm from a He-Ne laser and 532 nm from a solid-state laser. The laser plasma lines were removed using a BragGrate$^{\rm TM}$ bandpass filter (OptiGrate Corp.), and low-frequency Raman measurements down to 5 cm$^{-1}$ were achieved with three BragGrate$^{\rm TM}$ notch filters (OptiGrate Corp.). The typical spectral resolution is 0.34 cm$^{-1}$ per CCD pixel when using an 1800 lines/mm grating with 633 nm excitation. The laser power was kept below 1 mW to prevent sample heating.
	
	\subsection{Simulations}
	The local field generated by AuNCs was investigated using the three-dimensional finite-difference time-domain (3D-FDTD) simulation in Lumerical software. The incident light was modeled as a plane wave with $x$-polarization. The Bloch boundary condition in the $xy$-plane was applied to emulate infinite periodicity, and perfectly matched layers along the $z$-direction eliminated artificial boundary reflections. A critical uniform grid resolution of 0.1 nm$\times$0.1 nm$\times$0.1 nm was implemented to minimize numerical dispersion errors. The dielectric function of Au incorporated a modified Drude-Lorentz formula to account for nanoscale surface effects, and the optical response of graphene was governed by the Falkovsky model embedded in the software. The simulations extended to 1500 fs with dynamic convergence monitoring, ensuring field residuals stabilized within machine precision thresholds. The components of the local electric field were extracted through Fourier analysis of time-domain data at the characteristic wavelength of 633 nm.
	
	\section{Acknowledgments}
	We acknowledge the support from the National Key Research and Development Program of China (Grant No. 2023YFA1407000), the Strategic Priority Research Program of Chinese Academy of Sciences (CAS) (Grant No. XDB0460000), National Natural Science Foundation of China (Grants No. 12322401, No. 12127807, and No. 12393832), CAS Key Research Program of Frontier Sciences (Grant No. ZDBS-LY-SLH004), Beijing Nova Program (Grant No. 20230484301), Youth Innovation Promotion Association, Chinese Academy of Sciences (No. 2023125), Science Foundation of the Chinese Academy of Sciences (Grant No. JCPYJJ-22) CAS Project for Young Scientists in Basic Research (YSBR-026).
	
	\section{Data availability statements}
	The datasets generated during and/or analyzed during the current study are available from the corresponding author upon reasonable request.
	
	\section{Contributions}
	H.W. and M.L.L. contributed equally to this work. P.H.T. conceived and supervised the project. H.W. prepared the samples with assistance from L.S.C., Z.J.W., Y.F.Z., Z.Y.S., J.W., Y.M.S. and Z.M.W.. H.W. performed the measurements with support from Z.J.W, Y.F.Z, T.Y.Z. and X.L.L.. H.W., M.L.L. and P.H.T. analyzed the data. H.W., M.L.L. and P.H.T. developed the theoretical model with contributions from S.Y., X.W., and B.R.. H.W., M.L.L. and P.H.T. co-wrote the manuscript, with input and feedback from all authors.
	
	\section{Conflict of interest}
	The authors declare no competing interests.
	
	\section{Supplementary information}
	Supplementary information accompanies the manuscript on the Light: Science \& Applications website (http://www.nature.com/lsa)

\end{document}